\documentstyle[12pt]{article}

\csname @addtoreset\endcsname{equation}{section}

\def\bseq{\begin{subequation}}  
\def\eseq{\end{subequation}}
\def\bsea{\begin{subeqnarray}}  
\def\esea{\end{subeqnarray}}

\newcommand{\bbox}{\lower.2ex\hbox{$\Box$}}

\newcommand{\beq}{\begin{equation}}
\newcommand{\eeq}{\end{equation}}
\newcommand{\bea}{\begin{eqnarray}}
\newcommand{\eea}{\end{eqnarray}}
\newcommand{\ena}{\end{eqnarray}}

\renewcommand{\a}{\alpha}
\renewcommand{\b}{\beta}

\renewcommand{\d}{\delta}

\newcommand{\pa}{\partial}
\newcommand{\g}{\gamma}

\newcommand{\e}{\epsilon}
\newcommand{\z}{\zeta}

\renewcommand{\L}{\Lambda}
\newcommand{\m}{\mu}

\newcommand{\p}{\pi}

\newcommand{\s}{\sigma}

\newcommand{\Db}{\bar{D}}

\newcommand{\Sigmab}{\bar{\Sigma}}
\newcommand{\Phib}{\bar{\Phi}}

\newcommand{\ad}{{\dot{\alpha}}}
\newcommand{\bd}{{\dot{\beta}}}
\newcommand{\gd}{{\dot{\gamma}}}

\newcommand{\Del}{\nabla}
\newcommand{\Delb}{\bar{\nabla}}

\begin{document}

\begin{titlepage}
\begin{flushright} IFUM-599-FT\\ 
\end{flushright}
\vfill
\begin{center}
{\LARGE\bf The nonminimal scalar multiplet coupled \\
to supersymmetric Yang-Mills}\\
\vskip 27.mm  \large
{\bf   Silvia Penati$^1$  and  Daniela Zanon$^2$ } \\
\vfill
{\small
  Dipartimento di Fisica dell'Universit\`a di Milano and\\
INFN, Sezione di Milano, via Celoria 16,
I-20133 Milano, Italy}
\end{center}
\vfill

\begin{center}
{\bf ABSTRACT}
\end{center}
\begin{quote}
We consider the coupling of nonminimal scalar multiplets to supersymmetric
Yang-Mills in four dimensions and compute the one-loop
contribution to the low-energy effective action in the abelian 
sector. 
We show that the resulting theory realizes the dual
version of the corresponding one from $N=2$ supersymmetric Yang-Mills.  
\vfill      \hrule width 5.cm
\vskip 2.mm
{\small
\noindent $^1$e-mail: penati@mi.infn.it\\
\noindent $^2$e-mail: zanon@mi.infn.it}
\end{quote}
\begin{flushleft}
December 1997
\end{flushleft}
\end{titlepage}


Recently much attention has been given to the study of low-energy effective 
actions for supersymmetric gauge theories in four dimensions. These 
investigations
were boosted primarily by the Seiberg and Witten's exact construction
\cite{SW}
of the abelian low-energy effective action for the
$N=2$ Yang-Mills theory. In a $N=1$ superspace formalism such
a theory is described in terms of chiral superfields, in the adjoint 
representation of the gauge group, minimally coupled to the vector
gauge multiplet.

In this paper we consider the corresponding situation where the chiral 
superfields are replaced by complex linear superfields \cite{Gates},
\cite{superspace}. It is well
known that the on-shell spectrum of the minimal scalar, described
by a chiral superfield, and of the nonminimal one, described by 
a complex linear superfield, are the same, the two differing
only in their auxiliary field content. 
The kinetic action for a complex linear superfield $\Sigma$, 
$\Sigmab$, satisfying $\Db^2 \Sigma=0$, $D^2 \Sigmab=0$ is
\beq
S_{kin}=-\int d^4x~d^4\theta ~\Sigmab \Sigma
\label{kin}
\eeq
In terms of component fields \cite{Gates},
\cite{superspace}, it becomes  (the minus sign in (\ref{kin}) 
has been chosen so that it gives the correct sign for
the kinetic term of the component scalar field)
\beq
S=\int d^4x~[\bar{B}\Box B-\bar{\z}^\ad i\pa_{\a\ad} \z^\a
-\bar{H}H+\b^\a\rho_\a+\bar{\b}^\ad\bar{\rho}_\ad
+\bar{p}^{\a\ad}p_{\a\ad}]
\label{component}
\eeq
with physical components $B=\Sigma|$, $\z_\a=D_\a\Sigmab|$
like in the chiral superfield, but different auxiliary modes
$\rho_\a=D_\a\Sigma|$, $H=D^2\Sigma|$, 
$p_{\a\ad}=\Db_\ad D_\a \Sigma|$,
$\bar{\b}_\ad=\frac{1}{2}D^\a \Db_\ad D_\a \Sigma|$.

The coupling of $\Sigma$ to the Yang-Mills superfield $V$ is easily
accomplished by defining covariant derivatives and covariantly linear 
superfields with respect to the gauge fields. We use the conventions
in ref. \cite{superspace}.
Thus we consider $N=1$
superfields $V$, $\Sigma$ and $\Sigmab$ Lie-algebra valued in
the adjoint representation, $V=V^a T_a$, $\Sigma=\Sigma^a T_a$,
$\Sigmab=\Sigmab^a T_a$ with $[T_a,T_b]=i f^c_{~ab} T_c$ and
$tr T_a T_b= K\d_{ab}$, and we introduce covariant derivatives,
in vector representation, 
\bea
\Del_\a= e^{-\frac{V}{2}} D_\a e^{\frac{V}{2}} \qquad &~&
\qquad \Delb_\ad =e^{\frac{V}{2}} \Db_\ad e^{-\frac{V}{2}}
\nonumber\\
&~&~~~~~~~~~~\nonumber\\
\Del_{\a\ad}&=& -i \{\Del_\a,\Delb_\ad\}
\label{covderiv}
\eea
We define the Yang-Mills field-strength 
\beq
W_\a=\frac{i}{2}[\Delb^\ad,\{\Delb_\ad,\Del_\a\}]
\label{fieldstrength}
\eeq
and covariantly linear superfields $\Sigma_c$, $\Sigmab_c$ 
subject to the constraints 
$\Delb^2 \Sigma_c=0$, $\Del^2 \Sigmab_c=0$. In vector representation
they are given by
\beq
\Sigma_c =e^{\frac{V}{2}} \Sigma e^{-\frac{V}{2}} \qquad
\qquad \Sigmab_c =e^{-\frac{V}{2}} \Sigmab e^{\frac{V}{2}}
\label{sigmacov}
\eeq
The gauge invariant classical action is 
\bea
S&=& \frac{1}{\b^2}\left[ \int d^4x~ d^2\theta ~\frac{1}{4}tr(W^\a W_\a)
+\int d^4x~ d^2 \bar{\theta} ~\frac{1}{4}tr(\bar{W}^\ad \bar{W}_\ad)
\right.\nonumber\\
&~&~~~~~~~~~~\left.-\int d^4x~ d^2\theta~ d^2 \bar{\theta} ~tr(\bar{\Sigma}_c 
\Sigma_c
)\right]
\label{covaction}
\eea
In terms of the prepotential $V$ and of the complex linear
superfields $\Sigma$'s,
with gauge transformations
\bea
e^{V'}&=&e^{i\bar{\L}} e^V e^{-i\L}\qquad \qquad 
e^{-V'}=e^{i\L} e^{-V }e^{-i\bar{\L}}
\qquad \qquad \Db_\ad \L=0~~~~,~~~~ D_\a \bar{\L}=0
\nonumber\\
\Sigma'&=&e^{i\L} \Sigma e^{-i\L}\qquad \qquad 
~\Sigmab' =e^{i\bar{\L}}\Sigmab e^{-i\bar{\L}}
\eea
it can be rewritten as
\beq
S=\frac{1}{\b^2} \int d^4x~ d^2\theta~ d^2\bar{\theta}~tr\left[-\frac{1}{2}
(e^{-V}D^\a e^V)\Db^2(e^{-V}D_\a e^V)-e^{-V} \Sigmab e^V \Sigma
\right]
\label{actionprep}
\eeq

In addition to the explicit $N=1$ supersymmetry, the action in
(\ref{covaction}) possesses an invariance under a
second set of global supersymmetry transformations
\bea
&~&\d \Sigmab_c=-i\e^\a W_\a \qquad \qquad\qquad
\d \Sigma_c= i\bar{\e}^\ad \bar{W}_\ad \nonumber\\
&~&\d \Del_\a =\e_\a \Sigma_c \qquad \qquad\qquad ~~~~
\d \Delb_\ad =-\bar{\e}_\ad \Sigmab_c
\label{N2SS}
\eea
 The corresponding variations of the 
field-strengths can be obtained using the definitions in
(\ref{fieldstrength})
\beq
\d W_\a= -\bar{\e}^\ad \Del_{\a\ad} \Sigmab_c -\frac{i}{2}
\bar{\e}^\ad  \Del_\a \Delb_\ad \Sigmab_c
\qquad \qquad \d \bar{W}_\ad= \e^\a \Del_{\a \ad}\Sigma_c
+\frac{i}{2}
\e^\a \Delb_\ad \Del_\a \Sigma_c
\label{SSW}
\eeq
The supersymmetry algebra in (\ref{N2SS}) 
and (\ref{SSW}) closes on-shell. 
Indeed the commutators of two such transformations give
\bea
&~&[\d_1 ,\d_2] \Sigmab_c = (\e_2^\a \bar{\e}_1^\ad-\e_1^\a \bar{\e}_2^\ad)
(i\Del_{\a\ad}-\frac{1}{2}\Del_\a\Delb_\ad) \Sigmab_c \nonumber\\
&~&[\d_1 ,\d_2 ]\Sigma_c = (\bar{\e}_2^\ad \e_1^\a-\bar{\e}_1^\ad \e_2^\a)
(i\Del_{\a\ad}-\frac{1}{2}\Delb_\ad \Del_\a) \Sigma_c  \nonumber\\
&~&[\d_1 ,\d_2] W_\a = (\bar{\e}_2^\ad \e_1^\b -\bar{\e}_1^\ad\e_2^\b )
(i\Del_{\b\ad} W_\a+c_{\b\a}\Delb_\ad \Del^\g W_\g) \nonumber\\
&~&[\d_1 ,\d_2 ]\bar{W}_\ad = (\e_2^\a \bar{\e}_1^\bd -\e_1^\a \bar{\e}_2^\bd )
(i\Del_{\a\bd} \bar{W}_\ad +c_{\bd\ad}\Del_\a\Delb^\gd \bar{W}_\gd)
\label{SSclos}
\eea
and the closure of the algebra is implemented once the equations of
motion, $\Delb_\ad \Sigmab_c=0$,
$\Del_\a \Sigma_c=0$, $\Del^\a W_\a=0$, $\Delb^\ad \bar{W}_\ad=0$,
are imposed. The $N=1$ description in terms of
complex linear superfields and gauge multiplets is missing auxiliary 
degrees of freedom needed for the off-shell closure of the $N=2$ 
supersymmetry algebra.
This is in contradistinction with the 
corresponding situation for covariantly chiral superfields coupled to
Yang-Mills: there the second supersymmetry which realizes the $N=2$
invariance closes off-shell. This statement can be easily checked
considering the $N=2$ Yang-Mills action written in terms of $N=1$ 
superfields
\bea
S&=&\frac{1}{g^2} \left[ \int d^4x~ d^2\theta ~\frac{1}{4}
tr({\cal{W}}^\a {\cal{W}}_\a)
+\int d^4x~ d^2 \bar{\theta} ~\frac{1}{4}
tr(\bar{{\cal{W}}}^\ad \bar{{\cal{W}}}_\ad) \right.\nonumber\\
&~&~~~~~~~~~~\left.+\int d^4x~ d^2\theta~ d^2\bar{\theta} ~tr(\Phib_c \Phi_c
)\right]
\label{covactionYM}
\eea 
where $\Phi_c$, $\Phib_c$ are covariantly chiral, antichiral 
superfields respectively, and ${\cal{W}}_\a$ is the Yang-Mills
field-strength. The action in (\ref{covactionYM}) is
invariant under
\bea
&~&\d \Phi_c=-i\xi^\a {\cal{W}}_\a \qquad \qquad \qquad
\d \Phib_c= i\bar{\xi}^\ad \bar{{\cal{W}}}_\ad \nonumber\\
&~&\d \Del_\a =\xi_\a \Phib_c \qquad \qquad \qquad~~~~
\d \Delb_\ad =-\bar{\xi}_\ad \Phi_c
\label{N2SSYM}
\eea
 with corresponding transformations for the 
field-strengths 
\beq
\d {\cal{W}}_\a= i\xi_\a \Delb^2\Phib_c-\bar{\xi}^\ad
\Del_{\a\ad} \Phi_c
\qquad \qquad\d \bar{{\cal{W}}}_\ad= -i\bar{\xi}_\ad 
\Del^2 \Phi_c +\xi^\a \Del_{\a\ad} \Phib_c
\label{SSWYM}
\eeq
Using (\ref{N2SSYM}) and (\ref{SSWYM}) it is straightforward
to compute the commutator algebra and verify the
off-shell closure.

\vspace{0.8cm}
We now go back to the action in (\ref{actionprep}) and consider its
quantization. We shall restrict our attention to the $SU(2)$ case,
with $f_{abc}=\e_{abc}$ and $(T_a)_{kl}=i\e_{kal}$.
The aim is to perform the following one-loop computation:
we give a vacuum expectation value to the complex linear superfields
in an abelian direction
\beq
\Sigma=(0,0,\Sigma) \qquad \qquad \Sigmab=(0,0,\Sigmab)
\label{VEV}
\eeq
Then we compute all one-loop diagrams with external $\Sigma$,
$\Sigmab$ and no spinor nor space-time derivatives acting on them.
Obviously this amounts to the determination of the leading contribution 
to the low-energy effective action, exactly as in the corresponding
calculation for the $N=2$ supersymmetric Yang-Mills theory.

In order to proceed in the perturbative calculation, it is convenient 
to make a quantum-background splitting,
$\Sigma\rightarrow \Sigma_Q +\Sigma$, 
$\Sigmab\rightarrow \Sigmab_Q +\Sigmab$, so that $\Sigma$ and $\Sigmab$
are the external fields given in (\ref{VEV}), while $\Sigma_Q$, $\Sigmab_Q$
and $V$ are the quantum fields. Since one does not
know how to perform superspace functional differentiation and 
integration on complex linear superfields, a direct quantization 
is not available. Here we follow the approach proposed 
in some recent papers \cite{GVPZ}, \cite{PRVPZ}: the 
linearity constraints are solved in 
terms of unconstrained gauge spinor superfields $\s^\a$, $\bar{\s}^\ad$
whose field-strengths are $\Sigma_Q=\Db_\ad \bar{\s}^\ad$, $\Sigmab_Q=D_\a\s^\a$.
The gauge invariance introduced in this manner
needs gauge fixing; in fact an infinite set of
invariances arises and it leads to a tower of ghosts.
In ref. \cite{GVPZ} it was shown how to proceed systematically
in the quantization using the Batalin-Vilkovisky method. In ref. \cite{PRVPZ}
this formulation has been applied to the calculation of the one-loop
$\b$-function for the nonlinear $\s$-model defined in terms of nonminimal
scalar multiplets. It has been shown that choosing gauge-fixing functions
independent of the external background all the ghosts essentially
decouple; the calculation becomes manageable and no formal (infinite)
manipulations are necessary.

As compared to the nonlinear $\s$-model case, 
we have to face here the extra requirement
of maintaining the gauge invariance with respect to the Yang-Mills
fields. This is most easily done in a covariant approach: following
the gauge-fixing procedure described in detail in ref. \cite{GVPZ},
the only thing we need to do is to replace
the flat derivatives with the corresponding covariant ones given in
(\ref{covderiv}). This automatically leads to the coupling of the
infinite tower of ghosts to the Yang-Mills fields. While in general
this might be quite difficult to handle, no extra complications arise
for the calculation we have in mind. In fact it is immediate to realize that
these ghosts never couple to the external background $\Sigma$, $\Sigmab$
and then they do not contribute at one loop.

In ref. \cite{PRVPZ} the $<\s^\a \bar{\s}^\ad>$ propagator has been computed
and we refer the reader to that paper for the details of the calculation.
The final result is  
\bea
<\s_\a^a \bar{\s}_\ad^b>&=&\frac{\b^2}{ K} \d^{ab}\left[ 
- \frac{i \pa_{\a \dot \a}}{\Box}
+\frac{3(kk'_1)^2+4-2k^{'2}_1}{4(kk'_1)^2}
i \pa_{\a \dot \a} \frac{D^2 \bar D^2}{\Box^2}+\right.
\nonumber \\
&&\left.+
\frac{3 k^2 -2}{4 k^2} i \pa_{\a \dot \a} \frac{D_\b \bar D^2 D^\b}{\Box^2}+
\frac{2-k^2}{4 k^2} i \pa_{\a \bd} i \pa_{\b \ad}
\frac{D^\b  \Db^\bd}{\Box^2}\right]\d^{(4)}(\theta-\theta')
\label{sigmaprop}
\eea
where $k$ and $k'_1$ are gauge parameters. The gauge fixing for the $N=1$ vector 
multiplet is performed in standard manner: as in refs. \cite{WGR} and 
\cite{DGRUZ} we choose the supersymmetric Landau gauge which gives
the $V$-propagator
\beq
<V^aV^b>=-\frac{\b^2}{K} \d^{ab} \frac{D^\a\Db^2 D_\a}{\Box^2}
\d^{(4)}(\theta-\theta')
\label{Vprop}
\eeq
Again the chiral ghosts introduced by the quantization in the
Yang-Mills sector do not couple to the external fields and therefore
do not enter a one-loop calculation.

After quantum-background splitting, from the action in (\ref{actionprep})
we have as relevant interactions the cubic vertices
\beq
V^{(3)}=-\frac{1}{\b^2}~tr(\Sigmab[V,\Sigma_Q])- 
\frac{1}{\b^2}~tr([\Sigmab_Q,V]\Sigma)
\label{V3}
\eeq
 and the quartic vertex
\beq
V^{(4)}=-\frac{1}{2\b^2}~ tr(\Sigmab[V,[V,\Sigma]])
\label{V4}
\eeq
First we notice that, since the quantum fields $\s_\a$, $\bar{\s}_\ad$
always interact with the external background through their field-strengths
$\Sigma_Q$, $\Sigmab_Q$, instead of the propagator in (\ref{sigmaprop})
it is sufficient to use the much simpler expression \cite{PRVPZ}
\beq
<\Sigmab_Q^a \Sigma_Q^b>= D^\a<\s_\a^a \bar{\s}_\ad^b>\Db^\ad=
\frac{\b^2}{K} \d^{ab}\left[\frac{D^2\Db^2}{\Box}+
\frac{D_\a \Db^2 D^\a}{\Box}\right]\d^{(4)}(\theta-\theta')
\label{Sigmaprop}
\eeq
Then we find convenient to consider effective $V$-propagators
which contain the sum of all possible insertions of the quartic 
interaction in (\ref{V4}). The sum is performed taking 
advantage of the fact that, since we want to compute the leading 
contributions to the effective action, no derivative acts on the external
fields. For the spinor derivatives present in (\ref{Vprop}) we use the 
relation $(D^\a \Db^2 D_\a)^n=(-\Box)^{n-1}~D^\a \Db^2 D_\a$ 
and obtain
\beq
<<V^aV^b>>=-\frac{\b^2}{K}\left( \matrix{(\Box+\Sigmab\Sigma)^{-1}&
0& 0\cr
0&(\Box+\Sigmab\Sigma)^{-1}&0 \cr
0 &0&\Box^{-1} \cr}
\right)^{ab} \frac{D^\a \Db^2 D_\a}{\Box}\d^{(4)}(\theta-\theta')
\label{effectiveVprop}
\eeq
Now the one-loop $\Sigma\Sigmab$ low-energy effective action is
easily computed, summing over graphs with $n$ three-point vertices 
$\frac{-1}{\b^2}tr(\Sigmab[V,\Sigma_Q])$ and $n$ vertices
$\frac{-1}{\b^2}tr([\Sigmab_Q,V]\Sigma)$,  with $n$ propagators
$<<V^aV^b>>$ and $n$ propagators $<\Sigmab_Q^a \Sigma_Q^b>$.
The $D$-algebra is straightforward. We have spinor derivatives from the 
propagators in (\ref{Sigmaprop}) and (\ref{effectiveVprop}). Since they all 
 have to stay in the loop, we repeatedly use the identities
\beq
D^\a \Db^2 D_\a(D^2\Db^2+D_\b \Db^2 D^\b)=
D^\a \Db^2 D_\a(\Db^2 D^2+\Db_\bd D^2 \Db^\bd)=
\Box D^\a \Db^2 D_\a
\eeq
to reduce the number of $D$'s and $\Db$'s. As a final step the $D$-algebra
is completed by making use of 
\beq
\d^{(4)}(\theta-\theta')D^2\Db^2\d^{(4)}(\theta-\theta')=
\frac{1}{2}\d^{(4)}(\theta-\theta')D^\a \Db^2 D_\a\d^{(4)}(\theta-\theta')
=\d^{(4)}(\theta-\theta')
\eeq
The total sum of these contributions gives
\bea
\Gamma_{(1)}&=&\int \frac{d^4p}{(2\p)^4} d^4\theta~ \frac{2}{p^2}
\sum_{n=1}^{\infty}\frac{1}{n} \left(2\frac{-\Sigmab\Sigma}
{p^2-\Sigmab \Sigma} \right)^n \nonumber\\
&~&=
-\int \frac{d^4p}{(2\p)^4} d^4\theta~ \frac{2}{p^2}
\left[ln(1+\frac{\Sigma\Sigmab}{p^2})-
ln(1-\frac{\Sigma\Sigmab}{p^2})\right]
\label{1loopvertices}
\eea
In addition one has the contribution from the effective
$V$-propagator in (\ref{effectiveVprop}) with no extra
background insertions: it gives
\beq
\Gamma_{(2)}=-\int \frac{d^4p}{(2\p)^4} d^4\theta~ \frac{2}{p^2}
ln(1-\frac{\Sigma\Sigmab}{p^2})
\label{1loopV}
\eeq
Summing the terms in (\ref{1loopvertices}) and (\ref{1loopV})
one obtains
\beq
\Gamma=-\frac{2}{(4\p)^2}\int_0^{\infty} dp^2 ~d^4\theta~ 
ln(1+\frac{\Sigma\Sigmab}{p^2})
\eeq
The integral is divergent: it gives rise to a wavefunction
renormalization of the $\Sigma$'s plus a finite contribution
\beq
\Gamma_{eff}(\Sigmab,\Sigma)=
\frac{-2}{(4\p)^2} \Sigmab\Sigma ~ln\left(\frac{\Sigmab\Sigma}{\m^2}\right)
\label{effectiveaction}
\eeq
where $\m$ is a renormalization mass. 
The result in (\ref{effectiveaction}) represents the one-loop low-energy
contribution to the $\Sigmab \Sigma$ lagrangian, exactly as in 
the corresponding calculation in terms of chiral superfields.
Now we elaborate on this.

\vspace{0.5cm}
The equivalence of the descriptions of the scalar multiplet
by the complex linear superfield and by the chiral superfield
can be understood in terms of a duality transformation. One writes a 
first order action
\beq
S=\int d^4x~ d^4\theta~\left[\Phib \Phi +\Sigma\Phi +\Sigmab \Phib\right]
\eeq
with $\Sigma$ satisfying the constraint $\Db^2 \Sigma=0$,
(which implies $\Sigma= \Db_\ad \bar{\s}^\ad$),
and $\Phi$ unconstrained. Using the equations of motion
to eliminate $\Phi$, $\Phib$, one obtains the linear superfield
action. Eliminating instead $\Sigma$, $\Sigmab$ whose equations of 
motion impose the chirality constraint on $\Phi$ ($\Phib$), 
$\Db_\ad \Phi=0$, ($D_\a \Phib=0$),
one obtains the standard chiral superfield action. In view of this
observation our one-loop calculation has shown that this classical 
duality of the matter fields is maintained at the quantum level.
Since we have shown that both theories in (\ref{covaction}) and 
in (\ref{covactionYM}) are invariant under a second supersymmetry which mixes
the matter fields with the gauge fields, we would expect 
the duality in the matter sector,
$\Phi_c \rightarrow \Sigma_c$, to be accompanied by
the electro-magnetic duality, ${\cal{W}}_\a  \rightarrow W_\a $, 
in the vector sector. 

However things do not quite work so simply.
We can start for example with the
linear multiplet $\Sigma_c$, satisfying the constraint 
$\Delb^2 \Sigma_c=0$ (which implies $\Sigma_c=\Delb_\ad \bar{\s}^\ad$)
and the gauge field strength
$W_\a =\frac{i}{2}[\Delb^\ad,\{\Delb_\ad,\Del_\a\}]$, and consider
the first order action
\bea
S&=&\frac{1}{g^2}~tr\left[\int d^4x~d^4\theta~ \Phib_c \Phi_c 
+\frac{1}{4}\int d^4x~d^2\theta~
{\cal{W}}^\a {\cal{W}}_\a +\frac{1}{4}\int d^4x~d^2\bar{\theta}~
\bar{{\cal{W}}}^\ad \bar{{\cal{W}}}_\ad \right]\nonumber\\
&~&+tr\int d^4x~d^4\theta~ (\Phi_c \Sigma_c
+\Phib_c \Sigmab_c) +\frac{i}{2}tr\int d^4x~d^2\theta~
{\cal{W}}^\a W_\a \nonumber\\
&~&-\frac{i}{2}tr\int d^4x~d^2\bar{\theta}~
\bar{{\cal{W}}}^\ad \bar{W}_\ad
\label{firstorderaction}
\eea
with unconstrained superfields $\Phi_c$, $\Phib_c$, and chiral
spinors ${\cal{W}}_\a$ and $\bar{{\cal{W}}}_\ad$.

Now, functional integration over $\Phi_c$, $\Phib_c$,
${\cal{W}}_\a$ and $\bar{{\cal{W}}}_\ad$, leads to the action
\beq
S=g^2~tr\left[-\int d^4x~d^4\theta~ \Sigmab_c \Sigma_c 
+\frac{1}{4}\int d^4x~d^2\theta~
W^\a W_\a +\frac{1}{4}\int d^4x~d^2\bar{\theta}~
\bar{W}^\ad \bar{W}_\ad \right]
\label{dualaction}
\eeq
which exactly matches the one in eq. (\ref{covaction}) with
the identification $\b^2=1/g^2$.
On the other hand, using the equations of motion to eliminate the superfields
$\Sigma$, $\Sigmab$ and $V$, we obtain the chirality
constraints 
\beq
\Delb_\ad \Phi_c=0 \qquad \qquad \Del_\a \Phib_c=0
\eeq
but {\em not} the Bianchi identities
\beq
\Del^\a {\cal{W}}_\a +\Delb^\ad \bar{{\cal{W}}}_\ad=0
\label{Bianchi}
\eeq
Instead of eq. (\ref{Bianchi}) one obtains the equations of
motion
\beq
\Del^\a {\cal{W}}_\a =0 \qquad \qquad \qquad
\Delb^\ad \bar{{\cal{W}}}_\ad=0
\label {eqmotion}
\eeq
so that in a certain sense the standard $N=2$ Yang-Mills theory is 
recovered only on-shell. We believe that this is just a consequence of 
the fact that for the theory in (\ref{covaction}) the second 
supersymmetry is an on-shell invariance.

We notice that the first order action in (\ref{firstorderaction})
is appropriate for describing the above mentioned
duality equivalence in the abelian case. Indeed for the abelian
theory we delete the $tr$ operation and consider the linear
multiplet satisfying the constraint $\Db^2 \Sigma=0$ and
the gauge field strength $W_\a=i\Db^2 D_\a V$. Using the 
field equations to eliminate $\Sigma$, $\Sigmab$ and $V$,
one recovers the chirality conditions on $\Phi$ and $\Phib$
and the Bianchi identities on ${\cal{W}}_\a$, $\bar{{\cal{W}}}_\ad$,
so that the abelian $N=2$ Yang-Mills theory is
reconstructed.

In the nonabelian case one should try to obtain a formulation of
the theory in terms of complex linear superfields and gauge fields
directly in a $N=2$ description. Indeed
in ref. \cite{SW} the complete, nonperturbative information about 
the low-energy effective action was obtained starting from the $N=2$ 
superspace formulation of the theory in (\ref{covactionYM}). 
In $N=2$ superspace the action is given by the chiral integral 
of the prepotential, a holomorphic function 
${\cal{F}}(W)$ being $W$ the $N=2$ gauge superfield strength 
\cite{3old}. For the action in (\ref{dualaction}) such a formulation
is not known. In order to move into $N=2$ superspace, first one should
learn which are the auxiliary degrees of freedom not present 
in the $N=1$ formulation of the theory. Second one should find
a $N=2$ superfield which might accomodate  
a complex linear superfield and a Yang-Mills field-strength 
among its $N=1$ components.
These issues seem worth to pursue.

In a different but complementary perspective, 
it could be interesting to consider
models with mixed matter, i.e. models constructed in terms of chiral 
and complex linear superfields along the lines studied 
in \cite{Jimetal}, and analyze
their interaction with the Yang-Mills gauge multiplet. 

As a final comment we observe that in order to have these systems
completely under control it would be necessary
to understand how to treat the infinite tower of ghosts introduced
by the Batalin-Vilkovisky quantization of the complex linear superfield.
From the results presented here it is clear that the contribution
to the one-loop low-energy effective action of the gauge field could
be obtained using $N=2$ supersymmetry. This indicates that the 
final answer is simple, and therefore we only need an intelligent 
way to handle efficiently the ghost calculation.

\medskip
\section*{\small{\bf Acknowledgments}}
\noindent
We would like to thank M. Grisaru for useful discussions.
\noindent
This work was supported in part by MURST and by the European Commission 
TMR programme
ERBFMRX-CT96-0045, in which the authors are associated to the University 
of Torino.


\newpage
\begin{thebibliography}{999}
\bibitem{SW} N. Seiberg and E. Witten, Nucl. Phys. {\bf B426} (1994) 19.
\bibitem{Gates} S.J. Gates and W. Siegel, Nucl. Phys. {\bf B187} (1981) 389.
\bibitem{superspace} S.J. Gates, M.T. Grisaru, M. Ro\v{c}ek and W. Siegel,
{\em Superspace} (Addison-Wesley, 1983)
\bibitem{GVPZ} M.T. Grisaru, A. Van Proeyen, D. Zanon, Nucl. Phys. {\bf B502}
(1997) 345. 
\bibitem{PRVPZ} S. Penati, A. Refolli, A. Van Proeyen, D. Zanon,
{\em ``The nonminimal scalar multiplet: duality, sigma-model, beta-function''},
IFUM-579-FT, KUL-TF-97/26, hep--th/9710166.
\bibitem{WGR} B. de Wit, M.T. Grisaru, M. Ro\v{c}ek, Phys. Lett. {\bf B374}
(1996) 297. 
\bibitem{DGRUZ} A. De Giovanni, M.T. Grisaru, M. Ro\v{c}ek, R. von Unge, 
D. Zanon, Phys. Lett. {\bf B409} (1997) 251. 
\bibitem{3old} G. Sierra and P.K. Townsend, in "Supersymmetry and
Supergravity 1983", proc. XIXth Winter School, Karpacz, ed. B. Milewski
(World Scientific, 1983), p. 396;\\
B. de Wit, P.G. Lauwers, R. Philippe, S.-Q. Su and A. Van Proeyen,
Phys. Lett. {\bf 134B} (1984) 37;\\
S.J. Gates, Nucl. Phys. {\bf B238} (1984) 349
\bibitem{Jimetal} S.J. Gates, Phys. Lett. {\bf B365} (1996) 132, Nucl. Phys.
{\bf B485} (1997) 145; \\
S.J. Gates, M.T. Grisaru, M.E. Knutt-Wehlau, M. Ro\v{c}ek and O.A. Soloviev,
Phys. Lett. {\bf B396} (1997) 167. 
\end{thebibliography}
\end{document}